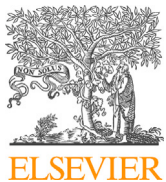
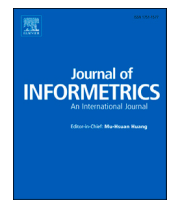

Research Paper

# Publishing instincts: An exploration-exploitation framework for studying academic publishing behavior and "Home Venues"

Teddy Lazebnik [a,b],[*],[1], Shir Aviv-Reuven [c], Ariel Rosenfeld [c,1]

[a] *Ariel University, Ariel, Israel*
[b] *University College London, London, UK*
[c] *Bar-Ilan University, Ramat-Gan, Israel*

A B S T R A C T

Scholarly communication is vital to scientific advancement, enabling the exchange of ideas and knowledge. When selecting publication venues, scholars consider various factors, such as journal relevance, reputation, outreach, and editorial standards and practices. However, some of these factors are inconspicuous or inconsistent across venues and individual publications. This study proposes that scholars' decision-making process can be conceptualized and explored through the biologically inspired exploration-exploitation (EE) framework, which posits that scholars balance between familiar and under-explored publication venues. Building on the EE framework, we introduce a grounded definition for "Home Venues" (HVs) – an informal concept used to describe the set of venues where a scholar consistently publishes – and investigate their emergence and key characteristics. Our analysis reveals that the publication patterns of roughly three-quarters of computer science scholars align with the expectations of the EE framework. For these scholars, HVs typically emerge and stabilize after approximately 15-20 publications. Additionally, scholars with higher h-indexes, greater number of publications, or higher academic age tend to have higher-ranking journals as their HVs.

## Introduction

Scholarly communication is the lifeblood of scientific progress, fostering the exchange of ideas, discoveries, and knowledge (Garvey William (1979); King et al. (2006)). Consequently, scholars' publishing behavior has attracted considerable attention in the academic community (Jamali et al. (2014); Abbott (2017); van Dalen (2021); Liu et al. (2023)). While the exact factors governing one's publication behavior may vary depending on the specific field and individual scholars' preferences, expectations, and goals, the venue's characteristics such as its aim and scope, relevance to the scholar's research topic and methodology, reputation, audience, editorial and reviewing practices, to name a few, often play key roles in shaping scholars' decisions (Shopovski and Marolov (2017); Carson et al. (2013); Calcagno et al. (2012); Kosyakov and Pislyakov (2024); Zhang et al. (2025)). Typically, many of these characteristics are readily available and can be assessed before publication; these include journals' key metrics (i.e., impact factor, subject classifications, rankings, etc.) (Gryncewicz and Sitarska-Buba (2021)), open access policies (Tennant et al. (2016); Shen et al. (2023)), format and publication frequency (Dwan et al. (2008)), and plenty more. Nevertheless, other factors such as the editorial and

---

reviewing quality and standards are inherently more subtle and can significantly vary between venues and even from one publication to the other within the same venue depending, for example, on the specific handling editor or set of reviewers assigned to a given submission (Orton et al. (2011)). This inherent, partially observed, variability brings about an additional layer of complexity to the scholar's decision-making process, which is significantly under-explored in comparison (van Lent et al. (2014); Kate et al. (2017)).

Existing literature in the field can be roughly divided into two groups: normative and descriptive. The normative studies aim to describe the "correct" or "desirable" decision-making process that scholars should perform. For instance, Salinas and Munch (2015) used a Markov decision process to propose two different submission decision-making processes - one to maximize citations and another that minimizes the number of resubmissions while taking into account acceptance probability, submission-to-decision times, and impact factors. The authors found that scholars should start with high-impact venues and "move down" in quality relatively quickly when balancing the two objectives. Similarly, Wong et al. (2017) assumed similar objectives while adopting a Pareto-front optimization with personalized preferences over the different objectives. The authors found that regardless of the time horizon of the optimization, scholars should submit to venues that maximize the impact factor times acceptance rate. Similarly, Adewumi and Popoola (2018) introduced a computationally effective model, based on a discrete multi-objective particle swarm optimization algorithm, for the submission process. A common limitation of normative works, such as the ones mentioned above, is that the explored models may not necessarily align with scholars' actual decisions, and typically focus on journal publications. In a complementary fashion, the studies belonging to the descriptive group aim to explore the decision-making process performed *de facto* by scholars. For example, Xu et al. (2023) identify four factors playing a role in the submission decision process to venues: factors related to information acquisition, factors related to journal evaluation, factors pertaining to submission outcome feedback, and factors related to the authors' backgrounds. The authors found that the first two factors have more impact on the submission process compared to the latter two. Dwan et al. (2008) highlighted the connection between reporting positive or significant results in a manuscript and its chances of being accepted to highly ranked venues due to review bias. The authors identified that this phenomenon often shapes scholars' decisions. Similarly, Mingers and Leydesdorff (2015) noted that scholars with stronger performance metrics tend to publish in venues with higher metrics as well. However, a common limitation of descriptive studies is that they do not provide a generative or mechanistic framework to explain and explore the observed phenomena. In other words, the underlying mechanisms driving scholars' publication choices remain unclear.

In this work, we aim to move beyond normative ideals and descriptive analyses of scholars' publication behavior by proposing a mechanistic framework for the observed fluctuation between frequent dissemination in a select few academic venues and occasional publication in a broader range of venues. Specifically, we propose the Exploration-Exploitation (EE) framework as an appealing candidate for modeling and exploring scholars' publication behavior (Mehlhorn et al., 2015). EE is a long-established conceptual framework that encapsulates the tension between exploring and exploiting options in the perusal of optimal outcomes (Trudel et al. (2021)). In fact, EE-based behaviors are highly common in nature and play a pivotal role in modeling and explaining many decision-making processes (Cook et al. (2013); Krongauz and Lazebnik (2023); Viseras et al. (2016); Cinotti et al. (2019); Lazebnik et al. (2024)). Most notably, various animal species face an EE trade-off when optimizing their chances of finding food, mates, and suitable habitats (Monk et al. (2018)). For instance, foraging birds exhibit exploration by scouting new locations to find food, while also exploiting reliable food sources they have come across in the past (Antoniou et al. (2013); Emlen (1952)). Similarly, certain fish species migrate to explore different environments during breeding seasons, while others stay in familiar habitats to exploit available resources consistently (Partridge (1982); Partridge et al. (1980)). The EE framework extends well beyond the realm of biology and generally applies to many human decision-making environments as well (Wilson et al. (2014); Berger-Tal et al. (2014); Sidhu et al. (2007); Rosenfeld and Kraus (2018)). For instance, companies explore multiple designs and prototypes before narrowing in on a small set of promising ones to exploit through manufacturing and marketing (Guo and Gershenson (2004); Dou et al. (2017)). Similarly, Lavie (2017) explored historical alliances, showing that nations often explore potential new alliances while affirming successful ones. See Fig. 1 for an illustration.

We argue that the EE framework offers a compelling set of concepts and models for the study of publication behaviors presented by scholars when navigating the vast landscape of academic venues. In particular, the EE framework can be instrumental in defining and investigating academic phenomena such as the formation of so-called "Home Venues" (HVs). That is, many scholars can typically name a (small) number of academic venues through which they frequently disseminate their work (Hyland (2016); Paasi (2005)). While there is no single agreed-upon term that captures this phenomenon, the elusive term HVs is often informally used to emphasize the idea that scholars tend to consider specific venues as their primary publishing platforms and intellectual homes within the academic community. These venues, which under the EE framework conceptually represent scholars' exploitation behaviors, are typically highly relevant to the scholar's field of study and attract the attention and communication of the desired audience. That is, by rather consistently submitting, and subsequently publishing, their work in these HV, scholars establish a presence within their scholarly community and effectively engage in valuable scientific communication (Li et al. (2008)). In fact, it was recently shown that high mobility across venues negatively correlates with citations, suggesting that successful scholars tend to publish consistently in a relatively small number of venues (Fan et al. (2024)). Nevertheless, from time to time, scholars may also publish their works in other venues for a variety of possible reasons, such as to explore different publication avenues, seek new feedback or recognition, and acclimate to evolving circumstances. These decisions, which under the EE framework conceptually represent exploration behaviors, allow scholars to gather information, uncover new opportunities, get fresh feedback and perspective on their work, extend their outreach, and ultimately, improve their academic status and future publication behaviors. Although the specific methods for balancing exploration and exploitation may vary across domains, decision-makers, and time, EE dynamics are expected to result in an uneven distribution, typically aligning with a power-law distribution and in particular, a Pareto distribution (Sun et al. (2018); Chen et al. (2009)). As such, in our context, one can presumably use a Pareto distribution fitting procedure to identify the (small) set of distinguishable venues through which a scholar disseminates a substantial portion of his/her work.

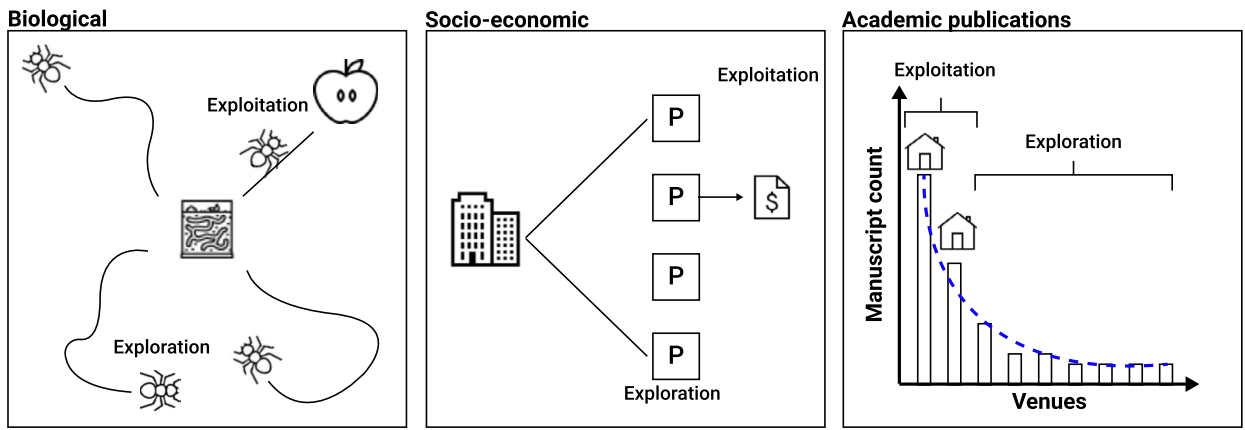

**Fig. 1.** An illustration of the EE framework applied to the biological, socio-economic, and academic publication settings.

It is important to note that the fundamental notion of uneven publication distribution, where a small subset accounts for a significant proportion of the overall distribution, has strong roots in the bibliometric literature as well. Notably, it strongly relates to Bradford's Law (also known as Bradford's Distribution or Bradford's Scatter) (Borgohain et al. (2021)), the 80-20 rule (also known as the Pareto Principle or the law of the vital few) (Burrell, 1985) and, more broadly, to the Pareto distribution (Nisonger, 2008). Intuitively, Bradford's Law states that if scientific venues are arranged in order of decreasing productivity of articles on a given subject, they can be divided into a nucleus of core ones, followed by a significantly increasing number of venues with a significantly decreasing number of publications relevant to that subject. The distribution is often assumed to follow a power-law pattern. A more general yet highly related concept is the 80-20 rule, a heuristic principle stating that roughly 80% of the effects come from 20% of the causes (Chen et al. (1993)). In our context, this rule would imply that 80% of a scholar's body of work is assumed to be published in 20% of the involved venues. The 80-20 rule can only be reflected precisely by the statistical distribution known as the Pareto distribution. In fact, the Pareto distribution has widespread applications, from modeling income inequality and customer behavior in business to analyzing seismic activity and species abundance in the natural world (Krasner (1991); Bierbrauer and Pierre (2014)). Given its versatility in modeling phenomena that exhibit significant inequalities or power-law relationships across diverse fields (including the study of science itself), as well as its ability to adequately model the observations of many EE dynamics, it is also adopted for defining HVs in this work. To our knowledge, little is currently known about the so-called "HV" phenomenon. In particular, a formal mathematical definition and the very existence of statistically distinguishable HVs have yet to be presented in the literature.

**Method**

*Formal model*

Let $s$ be a scholar with a body of work published in some set of venues, $J$. For each point in time ($t \in \mathbb{N}$), where $t$ is measured in years since the first publication made by $s$, we define $s$'s publication distribution, $H_t^s$, as a vector of size $|J|$ that contains simple counts of the number of publications made by $s$ up to time $t$ in each venue $j \in J$. For convenience, we assume $H_t^s$ is sorted from the largest to the smallest value with simple temporal tie-breaking (ordered by the time of inclusion in $J$), resulting in an ordered, non-increasing publication distribution. For simplicity, we denote $|J_t|$ as the number of unique venues scholar $s$ has published in up to time $t$.

For a given $s$ and $t$, $H_t^s$ may take one of multiple distributional forms. In order to differentiate between distributions that align with EE-like behavior and those which are not, we consider four standard, arguably archetypal and representative distributions, two of which consist of evident HVs (Panels A and B) and two that do not (Panels C and D), as illustrated in Fig. 2. Starting with Panel A, a single-peaked distribution (i.e., $f(x) = a$, where $x$ is a single venue and $a$ is the total number of publications of scholar $s$ at time $t$) may model a scholar who publishes all his/her work in a single venue. In this case, this venue should be considered a HV. However, since no "exploration" publications are present in this case, EE-like behavior seems highly unrealistic. In Panel B, a Pareto distribution ($\alpha \in \mathbb{R}$) may model a scholar who presents an uneven publication distribution that aligns well with EE dynamics. In this case, we define the scholar's HVs as any venue $j \in J$, such that its index in the sorted $H_t^s$ is smaller than the index of the "elbow point" in the fitted Pareto distribution (Bholowalia and Kumar (2014)). That is, following the common practice in the analysis of Pareto distributions (Krasner (1991)), one can first identify the point at which the cumulative distribution function (CDF) curve starts to exhibit a change in slope, signaling the transition from the power-law behavior to a more linear decline. Mathematically, this point can be obtained by minimizing the second derivative of the fitted Pareto distribution. All venues before the elbow point (i.e., on its left), often referred to as the "body" or "bulk" of the distribution, are deemed HVs. A venue with the largest number of publications in the HV set is termed the "leading" HV (ties are broken in favor of the venue with the earliest publication by the scholar). A formal definition of HVs is provided below. In the case of Panel C, a linearly declining distribution (i.e., $f(x) = -ax + b$ for $a, b \in \mathbb{N}$ where $x$ is the index of venue $x$ in the sorted $H_t^s$) may model a scholar who presents an uneven publication distribution that does not naturally align with EE-like dynamics, questioning the suitability of the EE framework. Similarly, in Panel D, a uniform distribution (i.e., $f(x) = \frac{1}{|J_t|}$ where $|J_t| > 1$ and $x$ is a venue in $H_t^s$) may model a scholar who publishes all his/her work evenly between various venues. In this case, it is evident that $s$ performed no "exploitation", making EE-like dynamics seem unrealistic as well. Presumably, in the case of linearly declining distribution and a uniform one, no obvious HV exists.

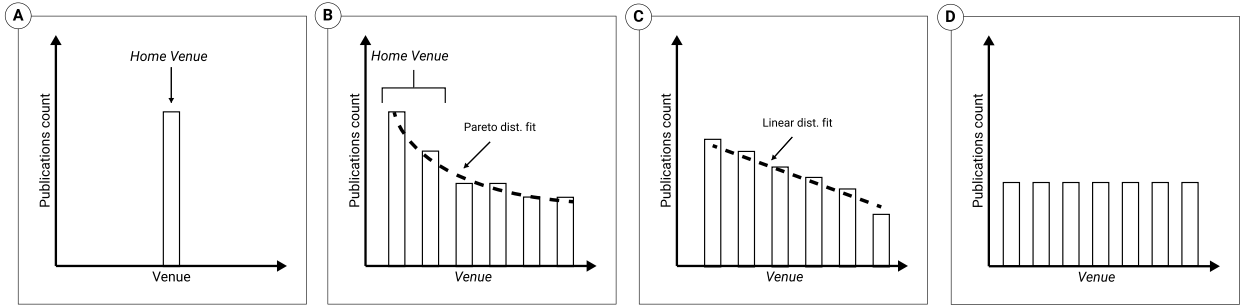

**Fig. 2.** A schematic view of the four publication distributions. (A) Presents a single-peaked publication distribution where all publications are disseminated through a single venue – i.e., the scholar's "home venue". (B) Presents an uneven publication distribution that roughly follows a Pareto distribution, indicating several "home venues". (C) Presents an uneven publication distribution roughly following a linear distribution, which does not indicate a clear "home venue". (D) Presents a uniform publication distribution where no "home venue" is evident.

Overall, a given publication distribution $H_t^s$ is deemed *aligned* with the EE-like behavior if it is better described by a Pareto distribution compared to single-peak, linearly declining, or uniform distributions. That is, we consider $s$ to have EE-aligned HVs at time $t$ if and only if the Pareto distribution provides the greatest adjusted coefficient of determination ($R^2$) out of the examined distributions (Ramberg et al. (1979); Bause (2020)).

Formally, assuming the publication distribution $H_t^s$ of scholar $s$ at time $t$ is best described by a Pareto distribution with parameter $\alpha$ (i.e., $f(x) := \alpha x_m^\alpha / x^{\alpha+1}$), we define the set of home venues $HV_t^s$ as follows: Let $k^*$ be the index of the elbow point in the sorted $H_t^s$, which is determined by minimizing the second derivative of the fitted Pareto distribution. That is,

$$k^* := argmin_{k \in [1,2,\ldots |J|]} |\frac{d^2}{dt^2} F(H_t^s[k], \alpha)|,$$

where $|J|$ is the number of unique venues scholar $s$ has published in up to time $t$, $H_t^s[k]$ is the number of publications in venue at index $k$ in the sorted publication distribution $H_t^s$, and $F(H_t^s, \alpha)$ is the cumulative distribution function of $H_t^s$ assuming a Pareto distribution with parameter $\alpha$. Then,

$$HV_t^s := \{j \in J | index(j, H_t^s) \leq k^*\},$$

where $index(j, H_t^s)$ denotes the index of venue $j$ in the sorted publication distribution $H_t^s$.

Furthermore, we define the leading home venue $j_L$ as:

$$j_L := argmax_{j \in HV_s^t} H_t^s[j] \tag{1}$$

with ties broken by the earliest publication in the venue $j$.

*Analytical approach*

We apply a three-step analysis workflow, as presented in Fig. 3. First, we retrieve and cross-reference contextual data from DBLP, Scopus, and Journal Citation Reports (JCR). Then, we apply our formal model to the data and analyze the publication distributions and their associated HVs (when applicable). Finally, we explore the underlying dynamics of scholars' decision-making and HVs through cluster-based analysis.

*Data preparation*

We consider the DBLP database, a specialized bibliographic database that provides open bibliographic information on major computational venues with decent coverage and accuracy over Computer Science (CS) literature (Rosenfeld (2023)). The DBLP database has been widely used in scientometric research and is often deemed as well-representative of the CS community (Lazebnik and Rosenfeld (2023); Kim (2018); Biryukov and Dong (2010); Kim (2019b); Cavacini (2015)). We retrieved the complete DBLP dataset on January 20th, 2024. At this point in time, the dataset contained roughly 21.4 million publications from almost 2.6 million scholars which were published in over 16 thousand venues. DBLP indexes several types of publications, most notably journal articles, conference papers, and informal publications. Together, these types cover roughly 90% of all indexed publications. For our analysis, we omitted informal publications, such as arXiv preprints (these are clearly marked in DBLP's data), as these are not peer-reviewed works and are only recently sufficiently covered by DBLP[2] (Aviv-Reuven et al. (2024); Santos et al. (2022)). Aligned with prior work (Aviv-Reuven and Rosenfeld (2021); Alexi et al. (2024)), we further exclude scholars with less than 5 publications. Overall, our resulting data consists of 0.65 million scholars and 18.3 million publications ($28.180 \pm 111.607$) publications per scholar.

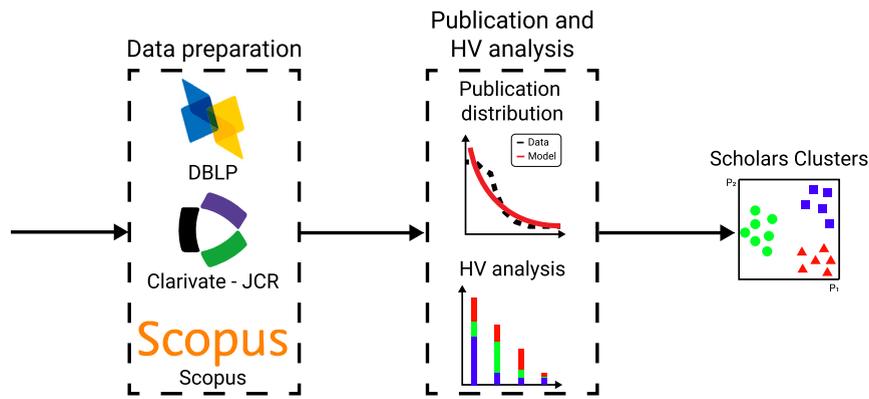

**Fig. 3.** A schematic view of the study's three-step workflow.

The data from DBLP was then cross-referenced with data from Scopus and JCR. Specifically, scholars' H-indexes, number of publications, publication years, and journals' quartile rankings (denoted Q ranking), as of 2022, were retrieved and matched accordingly. In order to reduce author and journal name ambiguity, when multiple records were available, we manually matched the records based on the available information one by one. Overall, we successfully matched 292,158 scholars (91.5%) and 891 journals, whereas the rest were removed from further consideration to avoid under-quality data. Given the ongoing debate on the ranking and evaluation of other publication types (i.e., conferences, books, etc) (van der Aalst et al. (2023); Küngas et al. (2013); Li et al. (2018)), we did not assign a counterpart value to the Q ranking for non-journal publications. In cases where a journal is ranked under multiple categories, we used the highest quartile available among the CS categories.

*Publication and HV analysis*

Algorithm 1 outlines the central computation process. Let $curve\_fit$ be a fitting procedure that receives a publication distribution and an assumed statistical distribution and returns the parameters associated with the optimal fit and its resulting coefficient of determination ($R^2$) (in our case, we use the least mean squares method (Björck (1996))) and later contrast it with the Maximum Likelihood Estimation (MLE) method (Pan et al., 2002). Formally, given a model function $f_p(x)$, where $x$ represents the independent variable(s) and $p$ is a vector of parameters, and given a set of data points $(x_i, y_i)$, $curve\_fit$ aims to find the parameter vector $p^*$ that minimizes the sum of the squared residuals between the observed values $y_i$ and the model's predicted values $f_p(x_i)$ which can be formalized as $p^* = argmax_p \sum_{i=1}^{n} (y_i - f_p(x_i))$. As we choose to use the least mean squares approach, we adopted the Levenberg-Marquardt algorithm (Moré, 2006). In addition, let $diff$ be a function that computes the numerical differentiation of a given vector (in our case, we use the forward Euler scheme (Cullum (1971))).

---

**Algorithm 1** Identifying Home Venues.

1: **Input:** $H_t^s$
2: **Output:** Number of HVs ($h$)
3: $dist_{sp}, R_{sp}^2 \leftarrow curve\_fit(H_t^s, \text{"Single-Peak"})$
4: $dist_p, R_p^2 \leftarrow curve\_fit(H_t^s, \text{"Pareto"})$
5: $dist_l, R_l^2 \leftarrow curve\_fit(H_t^s, \text{"Linear"})$
6: $dist_u, R_u^2 \leftarrow curve\_fit(H_t^s, \text{"Uniform"})$
7: $Best = argmax\{R_{sp}^2, R_p^2, R_l^2, R_u^2\}$
8: **if** $Best =' Single-Peak'$ **then**
9:   **Return** the single venue of $H_t^s$
10: **end if**
11: **if** $Best =' Pareto'$ **then**
12:   $h \leftarrow argmin(diff(dist_p, 2))$
13: **else**
14:   $h \leftarrow 0$
15: **end if**
16: **Return** first $h$ venues in $H_t^s$

---

First, we apply Algorithm 1 to each scholar in the dataset considering his/her entire publication distribution. Then, for each scholar with a non-empty set of HVs, we explore his/her HVs and non-HVs characteristics. That is, we consider the partial publication distributions starting from the first five publications onwards and re-apply Algorithm 1 to identify the periods for which HVs exist. A given publication is deemed an HVs "emergence" point if the scholar's HVs set becomes non-empty as the result of that publication and remains non-empty thereafter. Second, we explore the number, types, and rankings of HVs with and without considering their scholars' H-index, number of publications and academic age, using standard statistical testing. We further compare the Q rankings of scholars' HVs to their non-HVs, considering the key characteristics of scholars for whom non-HVs are ranked higher, equally, or lower than their HVs.

---

[2] https://dblp.org/statistics/recordsindblp.html.

**Table 1**
Scholars' publication distributions alignment with Pareto and non-Pareto distributions.

| Fitting | Portion | $R^2$ |
| --- | --- | --- |
| Pareto | 73.58% | $0.93 \pm 0.07$ |
| Non-Pareto | 26.42% | $0.40 \pm 0.14$ |

*Cluster-based analysis*

Considering scholars with an HV "emergence" point, a time series is extracted starting at the emergence point onwards consisting of the $\alpha$ parameter associated with the Pareto distribution after each publication. Conceptually, this time series represents the dynamics of a scholar's decision-making by describing how the balance between exploration and exploitation varies or stabilizes over time. Since scholars differ in their number of publications following an emergence point, we consider only those with at least 30 data points (i.e., at least 30 publications following the emergence point) and focus on the first 30 data points in our analysis. We cluster the resulting time series using a standard time series k-means algorithm (Tavenard et al. (2020)), finding the desirable number of clusters ($k$) using the elbow point method (Bholowalia and Kumar (2014)).

**Results**

We start by examining the publication distribution associated with each scholar in the data. Table 1 presents the portion of scholars for whom their publication distribution best aligns with a Pareto distribution. The results show that almost three out of four scholars (73.58%) are best aligned with a Pareto distribution and that, in these cases, the Pareto distribution seems to explain well the variability in the data ($R^2$ of $0.93 \pm 0.07$). In comparison, for the remaining scholars (26.42%), the variability in the data seems to be inadequately explained by the three alternative distributions ($R^2$ of $0.40 \pm 0.14$). The difference in $R^2$s between the two groups is statistically significant using a two-tailed t-test at $p < 0.05$.

For illustrative purposes, Fig. 4 demonstrates the application of Algorithm 1 to the publication distribution of the first and last authors of this article.[3] The results seem to exhibit the same pattern observed in Table 1, i.e., the first author's publication distribution best aligns with a linear distribution (albeit to a very limited extent – $R^2 = 0.214$), whereas the last author's publication distribution seems to fit well with a Pareto distribution ($R^2 = 0.964$).

Focusing on the scholars with a non-empty set of HVs, the results show that virtually all of them are best aligned with a Pareto distribution (99.97%). Furthermore, most of these scholars have an emergence point where all partial publication distributions following that point are best aligned with a Pareto distribution (93.62%). In other words, very few scholars align with a single-peaked distribution and once a Pareto distribution is found to bring about the best fit at a given point in one's career, the Pareto distribution generally remains the best fit from that point onwards. Given the high prevalence of this scholar population, we focus on them in the ensuing analysis: Fig. 5 shows the portion of scholars for whom a Pareto distribution brings about the best fit as a function of the number of publications (X-axis). By fitting a sigmoid function to the data, we obtained that $\rho = 0.75/(1 + e^{-0.46(x-12.98)})$ brings about the optimal sigmoid fit with a high coefficient of determination of $R^2 = 0.968$, which suggests that a sigmoid relation can well describe the data. Note that Fig. 5 is truncated at the 25th publication for presentation clarity only as the Sigmoid-like pattern remains relatively stable thereafter.

Considering these scholars, their average H-index, number of publications, and academic age are $14.27 \pm 15.71$, $46.28 \pm 200.02$, and $25.5 \pm 16.22$, respectively. The medians of these metrics are 9, 21, and 23, respectively. Fig. 6 shows the distribution over the number and types of HVs. Specifically, for each number of HVs (ranging from 1 to 4), the figure depicts the distribution over the types of HVs present in the HV set. We consider four cases: journals only (i.e., all HVs are journals), conferences only (i.e., all HVs are conferences), other (all HVs belong to the same type yet they are not journals or conferences), and a mixed case (different types). For clarity of presentation, we do not depict a very small number of scholars (less than 0.01%) for whom more than four HVs exist. Clearly, no scholars belong to the mixed case among those with only a single HV. Similarly, as the number of HVs increases, the mixed group becomes more prominent. Overall, 115,694 scholars (39.6%) have only journals as their HVs, 111,977 scholars (38.3%) only conferences as their HVs, 20,270 scholars (6.9%) have a different type of HVs, and 44,217 scholars (15.1%) have a combination of HV types.

---

[3] The second author, who is not presented in the figure, is a PhD student at the time these lines are written and has yet to acquire the necessary five DBLP-indexed publications for inclusion.

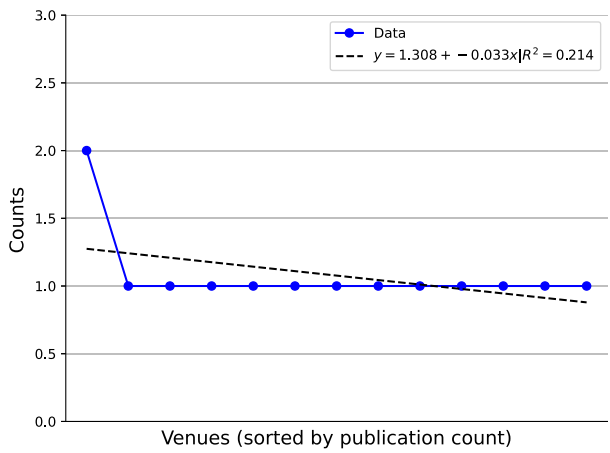
(a) First author.

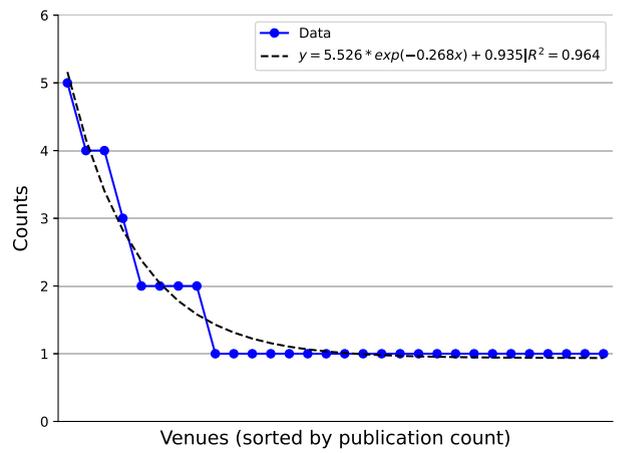
(b) Last author.

**Fig. 4.** An illustration of the application of Algorithm 1 to the publication distributions of the first and last authors of this article.

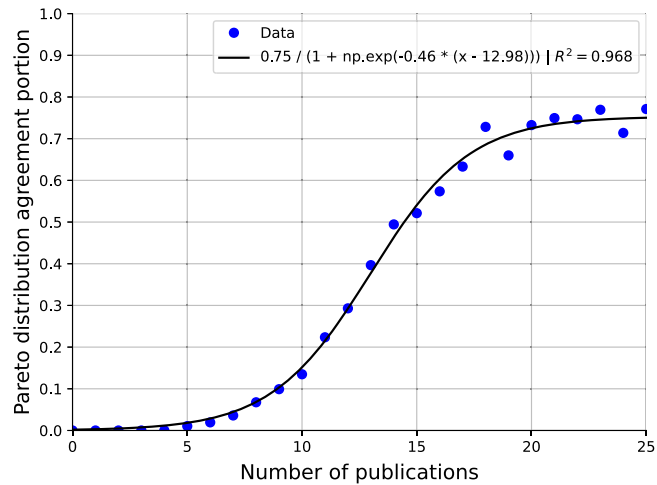

**Fig. 5.** The portion of scholars whose publication distribution best aligns with a Pareto distribution as a function of the number of publications.

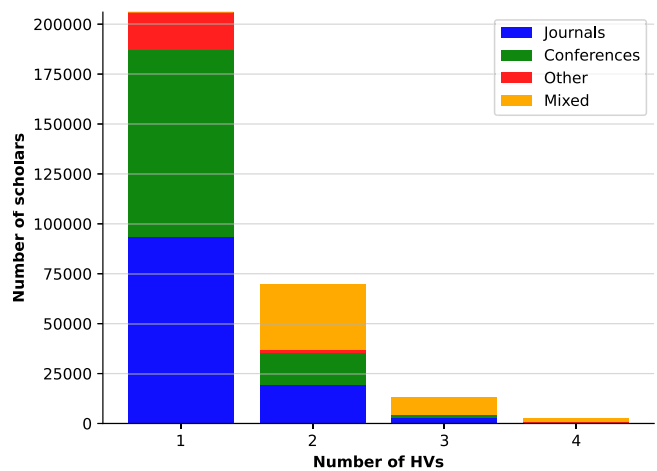

**Fig. 6.** Distribution of number and types of HVs.

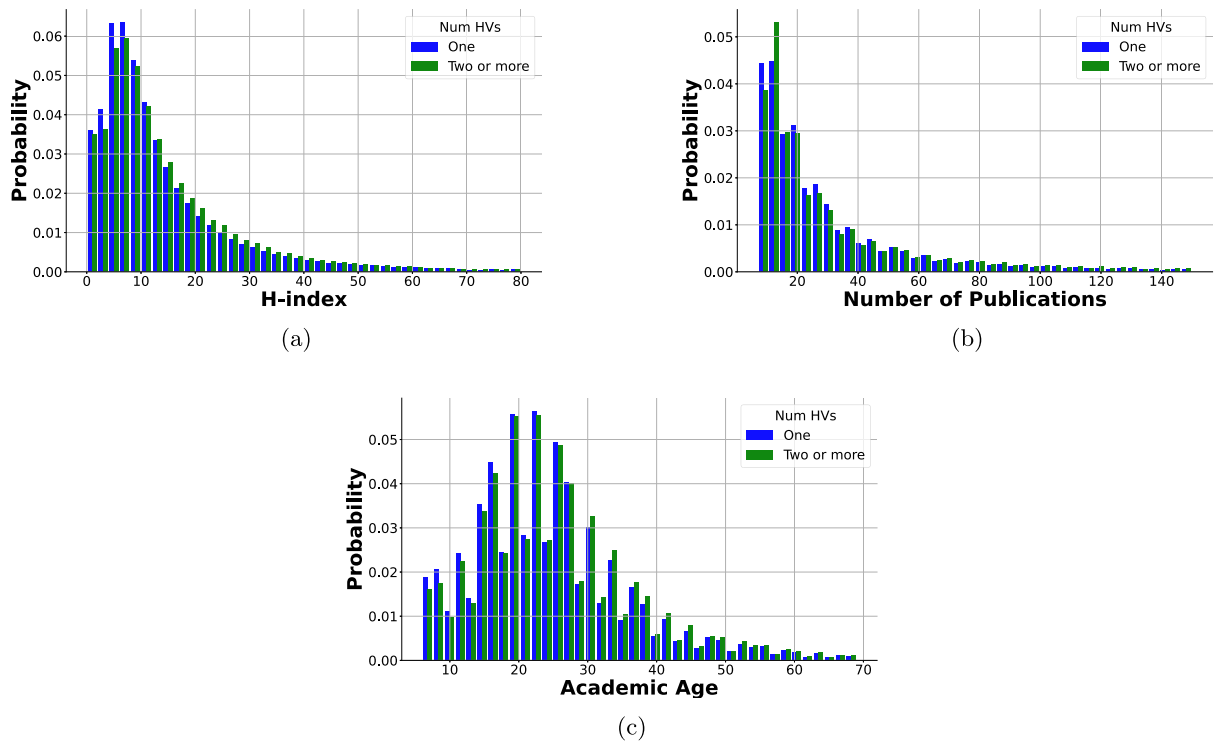

**Fig. 7.** H-index, number of publications and academic age of scholars with a single vs multiple HVs.

As the majority of scholars in the data have a single HV, we split the scholars into two groups: those with a single HV and those with multiple HVs. Figs. 7a, 7b and 7c display their distributions of H-indexes, number of publications and academic age, respectively. The mean (median) H-index of scholars with a single HV is 13.16 (9) compared to 14.07 (10) of scholars with multiple HVs. The mean (median) number of publications for scholars with a single HV is 34.6 (21), compared to 37.43 (21) for those with multiple HVs. The mean (median) academic age for scholars with a single HV is 24.57 (23), compared to 25.27 (23) for those with multiple HVs. The differences are statistically significant using a Mann-Whitney test (McKnight and Najab (2010)) at $p < 0.0001$.

In order to get a more nuanced perspective on the matter, Figs. 8a and 8b present the distributions of H-indexes according to the type of HVs as well as their number (i.e., a single or multiple HVs). Considering scholars with a single HV, those with a journal as their single HV have a mean (median) H-index of 14.57 (10) compared to those with a conference HV or other HV who have a mean (median) H-index of 12.26 (9) and 10.68 (7), respectively. A similar pattern emerges for scholars with multiple HVs where those with HVs comprised solely of journals have a mean (median) H-index of 16.07 (12) while those with only conferences and only other types of venues have 13.00 (9) and 9.45 (7), respectively. Scholars with a mixture of HV types have a mean (median) H-index of 13.65 (10). Similarly, Figs. 8c and 8d show the distribution of the number of publications for scholars with a single HV compared to those with multiple HVs. Scholars with single HV differ in their number of publications such that scholars with a journal as their HV have a mean (median) of 32.59 (20) publications while those with a conference HV or other HV have a mean (median) of 37.3 (22) and 31.41 (20) publications, respectively. Similarly, for scholars with multiple HVs, the mean (median) number of publications is 33.48 (19), 41.57 (22), 27.39 (17), and 38.04 (21) for only journals, only conferences, only others, and mixed cases, respectively. Figs. 8e and 8f show the distribution of the academic age for scholars with a single HV compared to those with multiple HVs. Scholars with a journal as their HV have a mean (median) of 25.1 (23) academic age while those with a conference HV or other HV have a mean (median) of 24.2 (23) and 23.6 (22) academic age, respectively. Similarly, for scholars with multiple HVs, the mean (median) academic age is 25.5 (24), 25.06 (23), 23.56 (22), and 25.28 (23) for only journals, only conferences, only others, and mixed cases, respectively. Nearly all of the above differences are statistically significant using Kruskal Wallis testing followed by post-hoc pairwise testing with Bonferroni correction (McKight & Najab, 2010), at $p < 0.05$. The only three exceptions are the pairwise comparison of the number of publications for scholars with a single journal HV and those with a single other HV ($p = 0.3$), and the pairwise comparisons of both the h-index and academic age of scholars with all conference HVs vs scholars with mixed HVs ($p = 0.03$ and $p = 0.05$ respectively).

Focusing on journal HVs, we compare their Q rankings as observed in Table 2. It is important to note in this context that the distribution of Q rankings in the entire journal pool is uneven (first column), as explored in prior work (Aviv-Reuven & Rosenfeld, 2023). Using $\chi^2$ testing followed by pairwise post-hoc testing with Bonferroni correction (McHugh, 2013), the results point to statistically significant differences between the Q ranking distribution of the entire journal pool and those associated with single and lead journal HVs, both at $p < 0.0001$, while no significant difference is observed between the latter two distributions. Specifically, while the proportion of Q1-ranked journals in all three groups is roughly the same (43%-44%), Q2-ranked journals are particularly more prevalent as HVs (37%) compared to their reduced prevalence in the entire journal pool (22%). Consequently, Q3 and Q4-ranked are less frequent as HVs (Q3: 11%-12%, Q4: 8%) compared to their prevalence in the entire journal pool (Q3: 19%, Q4: 15%).

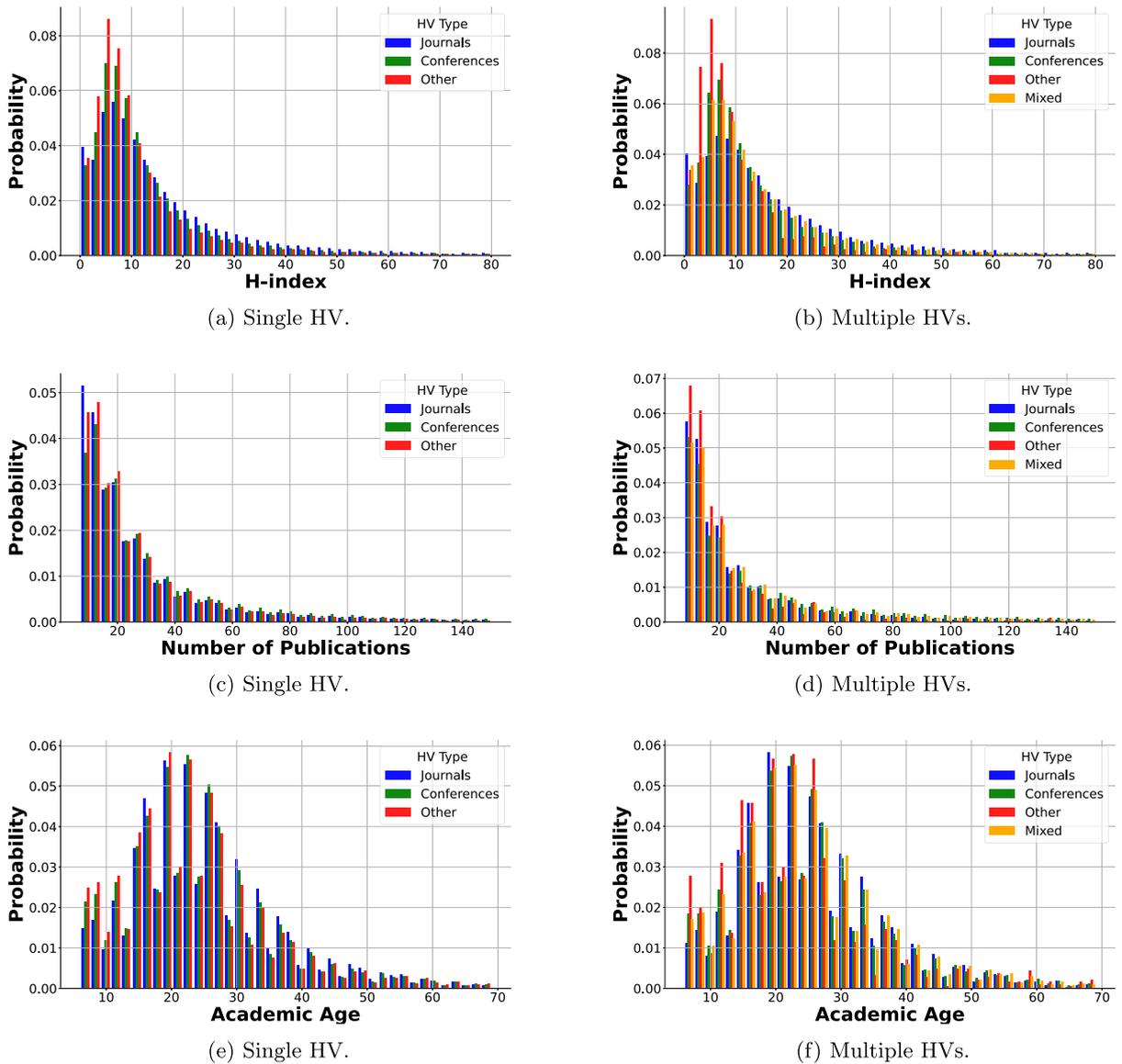

Fig. 8. H-index, number of publications, academic age and type of HVs based on the number of HVs.

**Table 2**
Journal ranking distributions across quarterlies (Q).

|    | Journal Pool | Single Journal HV | Leading Journal HV |
|----|--------------|-------------------|--------------------|
| Q1 | 0.44 (392)   | 0.44 (35687)      | 0.43 (11191)       |
| Q2 | 0.22 (194)   | 0.37 (30375)      | 0.37 (9522)        |
| Q3 | 0.19 (171)   | 0.11 (8822)       | 0.12 (3015)        |
| Q4 | 0.15 (134)   | 0.08 (7105)       | 0.08 (2158)        |

To get a more fine-grained perspective, we explore the possible relation between the H-index distribution of scholars and their single or leading journal HV's ranking distributions as presented in Figs. 9a and 9b. Notably, for both single and leading journal HVs, the mean, median, and H-index ranges seem to decrease as the Q ranking decreases. A Jonckheere trend test (Weller and Ryan (1998)) affirms that, indeed, scholars with lower-ranking single or leading journal HV tend to present lower H-indexes as well, at $p < 0.0001$. In a complementary fashion, Figs. 9c and 9d present the distribution over the number of publications compared to the single or leading journal HV's Q rankings. These figures do not seem to present any clear trend and, indeed, Jonckheere trend testing

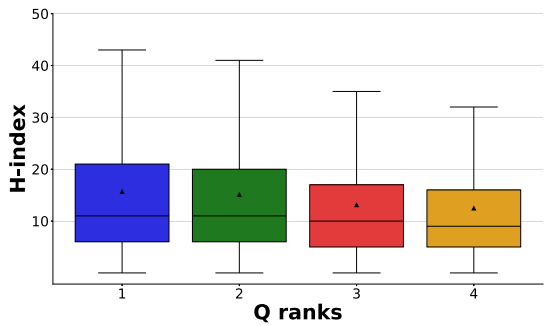
(a) H-index, single HV.

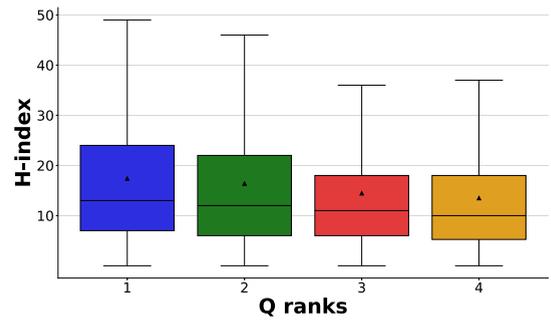
(b) H-index, multiple HVs.

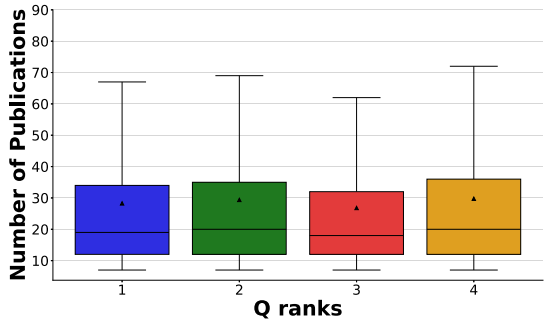
(c) Number of publications, single HV.

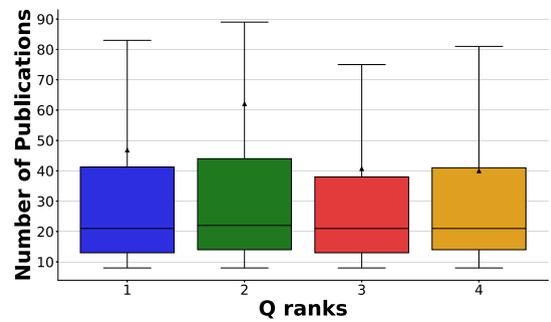
(d) Number of publications, multiple HVs.

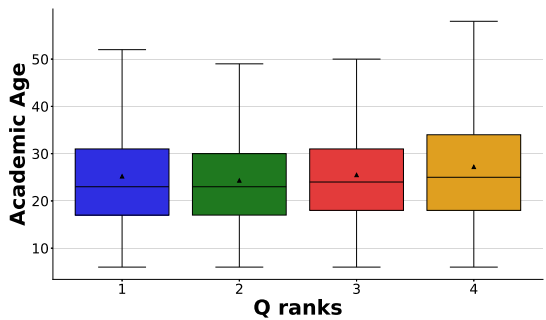
(e) Academic Age, single HV.

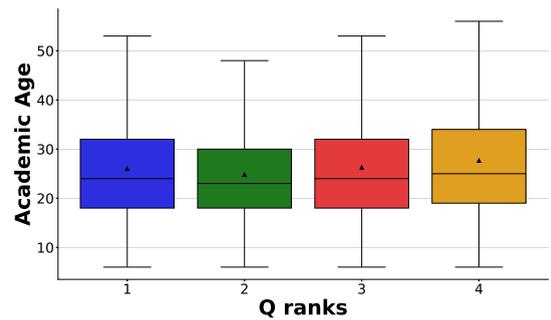
(f) Academic Age, multiple HVs.

**Fig. 9.** Distributions of scholars based on their number of HVs and journal rankings.

does not point to such a statistically significant trend. Similarly, Figs. 9e and 9f present the distribution over the scholars' academic age compared to the single or leading journal HV's Q rankings. As before, no statistically significant trend is found.

We also compare the median Q rankings of scholars' HVs and their non-HVs.[4] For this analysis, we focus on 131665 scholars for whom at least a single journal was present in both their HVs and non-HVs sets. We consider three groups: scholars for whom the median ranking of the HVs is higher than the median ranking of the non-HVs; scholars for whom the median ranking of the HVs is equal to the median ranking of the non-HVs; and scholars for whom the median ranking of the non-HVs is higher than the median ranking of the HVs. The case where the median Q rankings of a scholar's HVs and non-HVs are equal is the most prevalent (51664, 39.2%), followed by the case where the median Q ranking of a scholar's non-HVs is greater than that of HVs (41997, 31.9%) and the reverse (38004, 28.9%). We further consider the distribution of the H-index, number of publications, and academic age across the three groups as depicted in Figs. 10a, 10b, and 10c. Focusing on the H-index, scholars for whom the median ranking of the HVs is higher than the median ranking of the non-HVs have a mean (median) H-index of 15.39 (11) compared to scholars with an equal median ranking, and scholars with a higher non-HVs ranking who present an H-index of 15.25 (11) and 14.96 (11), respectively. For the number of publications, scholars with a higher median HVs ranking have a mean (median) of 28.8 (19) publications compared to scholars with an equal median ranking, and scholars with a higher non-HVs ranking who present 30.65 (21) and 30.38 (20) publications, respectively. For the academic age, scholars with a higher median HVs ranking have a mean (median) of 25.52 (24) years compared to scholars with an equal median ranking, and scholars with a higher non-HVs ranking who have a mean (median) academic age of 24.5 (23) and 25.66 (24), respectively. Using Kruskal Wallis with pairwise post-hoc testing with Bonferroni correction (McKight & Najab, 2010), nearly all of the above differences are statistically significant at $p < 0.05$. The only exceptions are the pairwise comparison of the h-index between scholars in the higher ranked non-HVs group and the equally ranked HVs and non-HVs group ($p = 0.46$), and the pairwise comparisons of the academic age between scholars in the higher ranked HVs group and those in higher ranked non-HVs group ($p = 0.08$).

---

[4] In cases where an even number of HVs are present, the HVs are ranked and the average of the two centered HVs is considered to be the median.

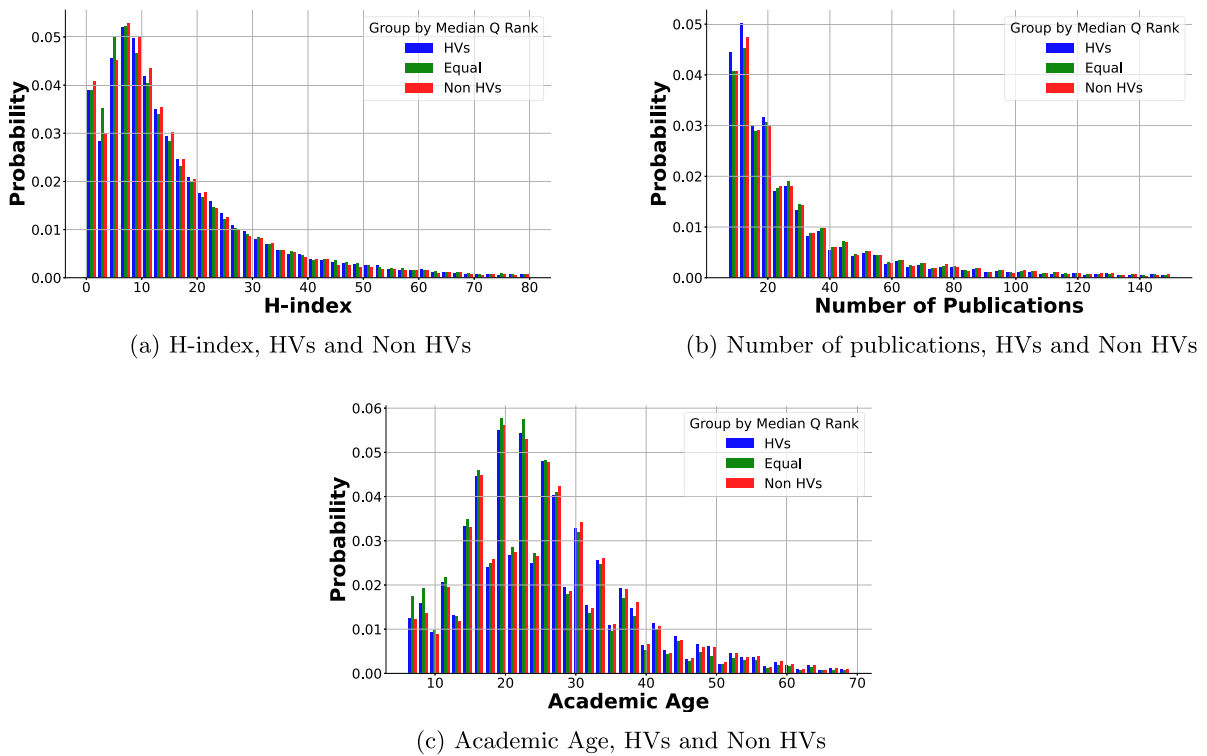

**Fig. 10.** H-index, number of publications, academic age and type of group, based on ranking of HV vs Non HV.

Finally, we perform a clustering-based analysis of the $\alpha$ parameter time series of scholars with an HV emergence point. As the number of publications varies across scholars, we normalized the time series by dividing the index of each publication in $H_t^s$ by the total number of publications by scholar $s$. Fig. 11a presents the three distinguished clusters identified through the elbow point method (Bholowalia and Kumar (2014)). These clusters seem to present three distinct patterns: stable (41.5%), decreasing (32.8%), and increasing (25.7%) $\alpha$. Starting with the "stable" pattern, these scholars seem to converge to a rather constant $\alpha$ value relatively quickly, indicating that their number of HVs is generally stable from that point onwards. The "increasing" pattern, on the other hand, points to a decreasing number of HVs over time (as the distribution becomes steeper). However, note that this pattern presents a noticeable "elbow" point from which the increase is very mild. Similarly, a "decreasing" pattern suggests that the number of HVs increases over time. In this pattern, the $\alpha$ values first increase rapidly but then start to decrease moderately over time.

In this context, it is important to note that highly celebrated scholars are present in all three groups. For example, Fig. 11b shows three Turing prize winners (arguably the highest award a CS scholar can receive): Judea Pearl, Jeffery Ullman, and Patrick Hanrahan which, according to our analysis, best align with the stable, decreasing, and increasing patterns, respectively.

**Discussion**

In this study, we propose and explore the use of the biologically inspired EE framework for studying the varying distributions of scholars' publications across different academic venues. Focusing on extensive data from the CS field, our analysis points to several intriguing outcomes.

Most notably, the results seem to support our hypothesis that, indeed, a significant portion (around three out of four) of CS scholars do present publication distributions that align well with those expected through the EE framework. In particular, as depicted in Table 1 and Fig. 5, the expected Pareto distributions, which seem to emerge for most scholars early on in their careers and persist thereafter, bring about significantly more favorable alignment with the data compared to other reasonable and common candidate models that do not naturally coincide with the EE framework. Taken jointly, we consider these outcomes to be substantial support for the plausibility of the EE framework to conceptualize and study scholars' publication decision-making and dynamics. In other words, our outcomes do not *validate* the EE-framework in our context but rather *support* its adequacy given the observed outcomes of the underlying decision-making process. While this is a common limitation of observational studies considering human decisions (Rosenfeld and Kraus (2018)), we plan to explore more detailed and specific EE-based mechanisms in the future, in order to further reinforce the EE framework adequacy and explanatory power. For example, the exact nature by which one balances exploration and exploitation may vary widely, and even change over time. A few example strategies include epsilon-greedy (Dann et al. (2022)) (i.e., choosing the most promising option observed thus far or performing a "random" action at a low probability) and more sophisticated alternatives (e.g., Softmax exploration, Thompson Sampling, Boltzmann exploration (Russo et al. (2018); Cesa-Bianchi et al. (2017))). It is important to note in this context that the adequacy of the EE-framework need not necessarily imply that a single strategy would be strictly adopted by scholars in practice. However, if one can point to prominent strategies that align with the observed behavior, such an outcome would further reinforce the appropriateness of the EE framework.

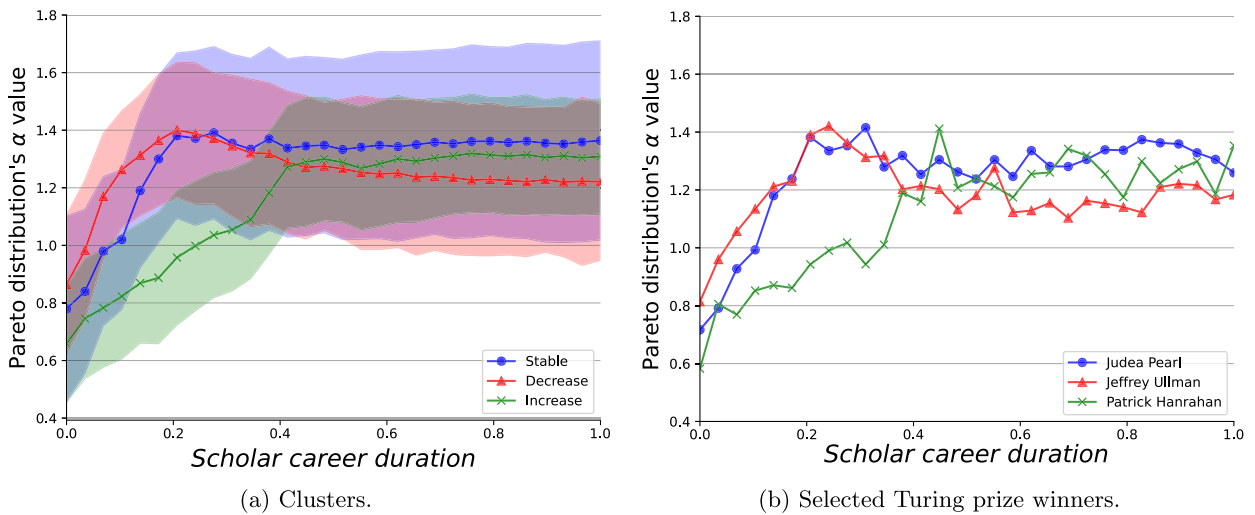

(a) Clusters.        (b) Selected Turing prize winners.

**Fig. 11.** Three temporal clusters obtained from the Pareto distribution's $\alpha$ value. The results are shown as mean ± standard deviation.

Using the EE framework, we mathematically define and explore the characteristics of the so-called HVs. As shown in Fig. 5, for the majority of CS scholars, HVs tend to emerge relatively early in one's career (after 15 to 20 publications). Most of these scholars seem to have a single HV, typically a journal or a conference. However, an exponential-like decrease in the number of scholars with multiple HVs is observed in Fig. 6, such that nearly all examined scholars have less than four HVs. These outcomes are generally aligned with the typical expectation in the CS field (as in various other fields (Kulczycki et al. (2018))) to publish mostly in journals and conferences (as opposed to books for example) (Kumari and Kumar (2020)) and the fact that most scholars can typically find more academic success by focusing on a small number of proximal venues (Lombardo et al. (2022)).

On average, scholars with higher H-indexes are more likely to have more than a single HV (see Fig. 7a), more likely to have an "all journals" HV set (see Figs. 8a, 8b) and, when their single or leading HV is a journal, that journal is typically higher ranked than those of other scholars 9a. These results are, presumably, not surprising. Prominent scholars are more likely to produce high-quality works that could be suitable for a variety of highly-ranked journals (Thelwall et al. (2023)). However, these results can be further explained by the traditional perspective on research evaluation that values journal over conference publications (Freyne et al., 2010). In particular, higher H-indexes, which are often considered a proxy of scholars' reputation (Thelwall & Kousha, 2021), may be bi-directionally associated with higher prevalence and ranking of journal HVs: since highly esteemed scholars tend to frequently publish in specific journals, these journals tend to receive higher attention and/or conversely, scholars who tend to publish in highly esteemed journals tend to receive more attention to their work (Klamer et al., 2002; Ding & Cronin, 2011). Further support for this explanation may be found in the association between academic age and one's likelihood to have a journal HV or "all journals" HV set as more seasoned scholars tend to hold more traditional perspectives on research evaluation (Milojevic, 2012).

While more productive scholars (i.e., more publications), on average, tend to have more than a single HV too (see Fig. 7b), they are more likely to have an "all conference" HVs (see Figs. 8c, 8d). When a single or leading journal HV is present, productivity does not seem to be associated with that journal's ranking. One reasonable explanation would be that CS conferences often offer a fast review and publication process, which need not necessarily indicate higher quality publications (Vrettas and Sanderson (2015)). It is also important to note that CS conferences are being increasingly recognized within the CS scientific discourse with an exponentially increasing publication volume in a growing number of highly related conferences (Kim (2019a)). With the increasing competition for positions, funding, and recognition (Wuchty et al. (2007); Von Bergen and Bressler (2017)), it is arguably not surprising that younger CS scholars tend to favor conferences over journal HVs.

The skewed distribution of journal HV rankings, as shown in Table 2, and the tendency of journal HVs to be ranked in Q2 significantly more often compared to their prevalence in the entire pool of CS journals (and significantly less often in Q3 and Q4), can be partially explained by the expected desire of most scholars to publish more frequently in higher-ranking journals. That is, while a scholar may, occasionally, publish in a low-ranking journal (i.e., Q3 or Q4), she is more likely to strive to build a more consistent presence in higher-ranking ones. Unfortunately, as consistently publishing at Q1-ranked journals is highly challenging (Kosyakov & Pislyakov, 2024), scholars' tendency to "resort" to Q2-ranked journals seems reasonable. In this context, it is important to note that scholars with higher H-indexes tend to be associated with higher-ranking (leading) journal HVs, further supporting the above explanation.

The comparison of journal HVs to journal non-HVs rankings seems to present no clear and consistent pattern. For example, while scholars with higher-ranked HVs tend to have higher h-indexes, scholars with equally ranked HVs and non-HVs and those with higher-ranked non-HVs tend to present greater productivity. These seemingly inconsistent observations can be partially explained by the stochastic and somewhat opportunistic nature of the exploration process that is assumed to govern non-HVs (Lai et al., 2010; Candelieri et al., 2022). For example, serendipity and unexpected discoveries that arrive from spontaneous observations or ideas rather than a direct pursuit of a research goal can result in outcomes being published outside one's usual community. Similarly, scholars who typically avoid open-access venues may be compelled to publish in them when utilizing grant funding that mandates open-access dissemination. Aligned with this explanation, scholars are also roughly equally divided into the three examined groups with only a minor tendency towards equal median ranking (39.2%), which can further attest to a stochastic exploration process. Having that said, both the higher ranked HVs and higher ranked non-HV groups were older, in terms of academic age, than the equally-rank group, suggesting that senior scholars may adopt more polarized publishing strategies, while equally-rank scholars tend to be earlier in their careers. This outcome aligns with our cluster-based analysis that revealed a non-negligible exploration-oriented pattern as well. Specifically, while scholars mainly follow the EE-like dynamics, the exact implementation thereof may change over time, as observed through the change in the $\alpha$ parameter over one's career (see reflected by Fig. 11). In particular, after the emergence of HVs, more than 40% of scholars seem to present a rather stable balance between exploration and exploitation as reflected by a stable pattern in their $\alpha$ values. Nevertheless, many scholars still seem to vary in their publication decisions, more toward exploration (32.8%, decreasing $\alpha$ values) than towards exploitation (25.7%, increasing $\alpha$ values). Interestingly, all three patterns seem to agree on an $\alpha$ value between 1.2 and 1.4 from some point onwards, suggesting some commonalities in their asymptotic behaviors. These results align with prior literature showing that more seasoned scholars tend to present less variability in their academic behavior (Zhang and Glänzel (2012); Rhodes (1983); Lazebnik and Rosenfeld (2023)).

Above all, our work proposes a biologically inspired and empirically grounded conceptual framework for studying scholars' publication decisions. However, we believe that the EE framework may also be suitable for other academic dynamics such as scholars' selection of research projects (i.e., balancing between familiar, low-risk projects and new, high-risk and potentially high-gain ones), collaboration patterns (i.e., balancing between established, reliable peers and new, potentially more innovative but uncertain collaborations), and mentorship and supervision (i.e., balancing the effort between nurturing established students with predictable progress and taking on new students with novel, but uncertain, potential), to name a few.

In this study, we adopted the Pareto distribution as a grounded instance of a power-law distribution that is often considered favorable in EE-based analysis (Nezami & Anahideh, 2024; Schütze et al., 2020; Ma et al., 2020). However, it is important to note that other, similar, power-law distributions may also align with the observed publication distributions. To verify the appropriateness of the Pareto distribution in our context, we further examine two common alternative power-law distributions: the exponential distribution (Balakrishnan, 1996) and the Frechet distribution (Afify et al., 2016), and replicate the analysis leading to Table 1. When considering each of these distributions as an alternative to the Pareto distribution, the results demonstrate only minor, non-significant differences, most of which are in favor of the Pareto distribution. Specifically, while the Pareto distribution best describes 73.53% of the population ($R^2 = 0.93 \pm 0.07$), the exponential distribution best describes 71.38% ($R^2 = 0.90 \pm 0.08$) and Frechet 74.2% ($0.92 \pm 0.05$). Nevertheless, when comparing the three power-law distributions head-to-head, the Pareto distribution brought the best fit in 62.7% of the cases, followed by the exponential distribution and Frechet, which scored 24.1%, and with 13.2%, respectively. Jointly, the Pareto distribution seems favorable in our setting as well.

A key limitation of our work is its reliance on observed publications rather than submissions. Clearly, non-accepted submissions are not publicly available and thus, cannot be readily obtained. To overcome this limitation, we launched a free online academic submission tracking system called Tracademic[5] that allows scholars of all disciplines and institutions to keep track of the status of their submissions. In addition to its potential applicative value to the scholars who use the service, the anonymized data collected in the process could be of great value for studying submission dynamics in the future. An additional possible limitation of our analysis is the use of least squares for distribution fitting, a technique that may not be optimal for precisely characterizing heavy-tailed distributions (Clauset et al., 2009). To address this potential limitation, we recomputed the results presented in Table 1 by replacing the least mean squares fitting method with the maximum likelihood estimation (MLE) method in the implementation of Algorithm 1. The results show only minor differences between the two approaches, with the MLE method classifying 0.34% more scholars into the Pareto group. However, these additional scholars are associated with a slight decline in $R^2$ scores among Pareto-classified scholars (0.90 vs. 0.93), and a modest increase among Non-Pareto scholars (0.45 vs. 0.40). A closer examination of the disagreement between the methods reveals that nearly all scholars classified as Pareto under the least squares method were also classified as such by the MLE (99.91%). Taken together, these results suggest that switching from least squares to MLE slightly extends Algorithm 1's ability to identify Pareto-aligned distributions. Yet, the few additional classifications appear to be poorly explained by the Pareto distribution itself. Intuitively, these cases may be considered "borderline Pareto distributions" and offer limited additional value to our subsequent analysis, which focuses on the dynamics and characteristics of clearly Pareto-derived HVs. Finally, it is important to note that the corresponding and first authors of a publication are typically considered to be the primary contributors to the publication (Mattsson et al., 2011; Lazebnik & Rosenfeld, 2025), as such, arguably, they may have a greater influence on the decision regarding the publication venue. A simple replication of the analysis leading to Table 1 and Fig. 5 using only first-authored publications has not pointed to statistically significant differences in the alignment with the Pareto distribution (all: 73.58% vs first-authored: 69.71%) and the coefficient of determination for the sigmoid fitting (all: 0.968 vs first-author: 0.923). Nonetheless, we intend to further explore the complex relationship between a scholar's role in a publication (e.g., student, lead, supervision, sole authorship), and his/her publication behavior in future work. Finally, it is important to note that our analysis focuses explicitly on CS literature as indexed by DBLP. To further generalize the reported outcomes, we intend to extend our analysis to include additional disciplines that need not necessarily align with the practices and standards of Computer Science (e.g., Humanities and Social Sciences) as well as the consideration of interdisciplinary and cross-disciplinary scholars.

---

[5] https://tracademic.info/.

**CRediT authorship contribution statement**

**Teddy Lazebnik:** Writing – review & editing, Writing – original draft, Visualization, Software, Project administration, Methodology, Investigation, Formal analysis, Data curation, Conceptualization. **Shir Aviv-Reuven:** Writing – original draft, Visualization, Software, Formal analysis, Data curation. **Ariel Rosenfeld:** Writing – review & editing, Writing – original draft, Validation, Supervision, Methodology, Investigation, Formal analysis, Conceptualization.

**Funding**

This research did not receive any specific grant from funding agencies in the public, commercial, or not-for-profit sectors.

**Declaration of competing interest**

The authors have no financial or proprietary interests in any material discussed in this article.

**Data availability**

The data that has been used is presented in the manuscript with the relevant sources.